\documentclass[num-refs]{WileyMSP-template}
\usepackage{graphicx}
\usepackage{color}
\usepackage{cite}

\begin{document}
\pagestyle{fancy}

\rhead{}

\title{Bulk Rashba spin splitting and Dirac surface state in $p$-type \\
(Bi$_{0.9}$Sb$_{0.1})_2$Se$_3$ single crystal}

\maketitle


\author{P. K. Ghose,} 
\author{S. Bandyopadhyay,} 
\author{T. K. Dalui,} 
\author{J.-C. Tseng,} 
\author{J. K. Dey,} 
\author{R. Tomar,}
\author{S. Chakraverty,} 
\author{S. Majumdar,} 
\author{I. Dasgupta$^{\ast}$,} 
\author{S. Giri$^{\dagger}$}


\dedication{}

\begin{affiliations}
P. K. Ghose, S. Bandyopadhyay, T. K. Dalui, J. K. Dey, Prof. S. Majumdar, Prof. I. Dasgupta$^{\ast}$, Prof. S. Giri$^{\dagger}$\\
School of Physical Sciences, Indian Association for the Cultivation of Science, Jadavpur, Kolkata 700032, India \\
Email Address:$\ast$ sspid@iacs.res.in\\
Email Address:$\dagger$ sspsg2@iacs.res.in\\

Dr.J.-C. Tseng\\
Deutsches Elektronen-Synchrotron DESY, Notkestr. 85, 22607 Hamburg, Germany \\

R. Tomar,Prof. S. Chakraverty\\
Nanoscale Physics and Device Laboratory, Institute of Nano Science and Technology, Phase-10, Sector-64, Mohali, Punjab 160062,India\\

\end{affiliations}


\begin{abstract}

We report bulk Rashba spin splitting (RSS) and associated  Dirac surface state in (Bi$_{0.9}$Sb$_{0.1})_2$Se$_3$, exhibiting  dominant $p$-type conductivity.
We argue from the synchrotron diffraction studies that origin of the bulk RSS is due to a structural transition to a non-centrosymmetric $R3m$ phase below $\sim$ 30 K. The Shubnikov-de Haas Van (SdH) oscillations observed in the magnetoresistance curves at low temperature and the Landau level fan diagram, as obtained from these oscillations, confirm the presence of nontrivial Dirac surface state. The magnetization data at low temperature exhibit substantial orbital magnetization consistent with the bulk RSS. The existance of both the bulk RSS and Dirac surface states are confirmed by first principles density functional theory calculations. Coexistence of orbital magnetism, bulk RSS, and  Dirac surface state is unique for $p$-type (Bi$_{0.9}$Sb$_{0.1})_2$Se$_3$, making it an ideal candidate for spintronic applications.

\end{abstract}


\section{Introduction}

Recently, topological insulators (TI) exhibiting Rashba spin-orbit splitting (RSS) due to the presence of 2D electron gas states have been recognized as key materials for next generation spintronic devices without the requirement of an external magnetic field for manipulation of spins \cite{man_rev,datta,nitt,hyun1,sante1}. The spin degeneracy in such nonmagnetic materials is lifted by the strong Rashba spin-orbit coupling (SOC) associated with the broken space inversion symmetry. Rashba effect was first reported in the bulk wurtzite crystals\cite{rashba} and subsequently it was realised in the two-dimensional (2D) electron gases \cite{bych,valentine1}. 
Bi$_2$Se$_3$, a well known 3D TI,\cite{hyd,daniel1} attracted special attention, when the features of 2D electron gas was proposed to coexist with the nontrivial surface state\cite{bian_BS0,analy1}. The RSS of the 2D electron gas in Bi$_2$Se$_3$ has been experimentally realized, suggesting electrostatic control of spin splitting necessary for spintronic applications without  magnetic field\cite{king_BS1}. Recently, doping in Bi$_2$Se$_3$ was found to promote RSS due to band bending induced realisation of 2D electron gas states. In particular, the gapless surface state and RSS was found for the Cr doping in Bi$_2$Se$_3$\cite{wang_BS3}. The tunable Rashba-like spin-polarized state was also suggested for K-doping in Bi$_2$Se$_3$\cite{zhu_BS2}. Recently discovery of unusual orbital magnetism in Bi-rich Bi$_2$Se$_3$ nanoplatelets\cite{kim_BS5} was attributed to RSS and correlated with $n$-type conductivity of this system.

\begin{figure}
\includegraphics[width = \linewidth]{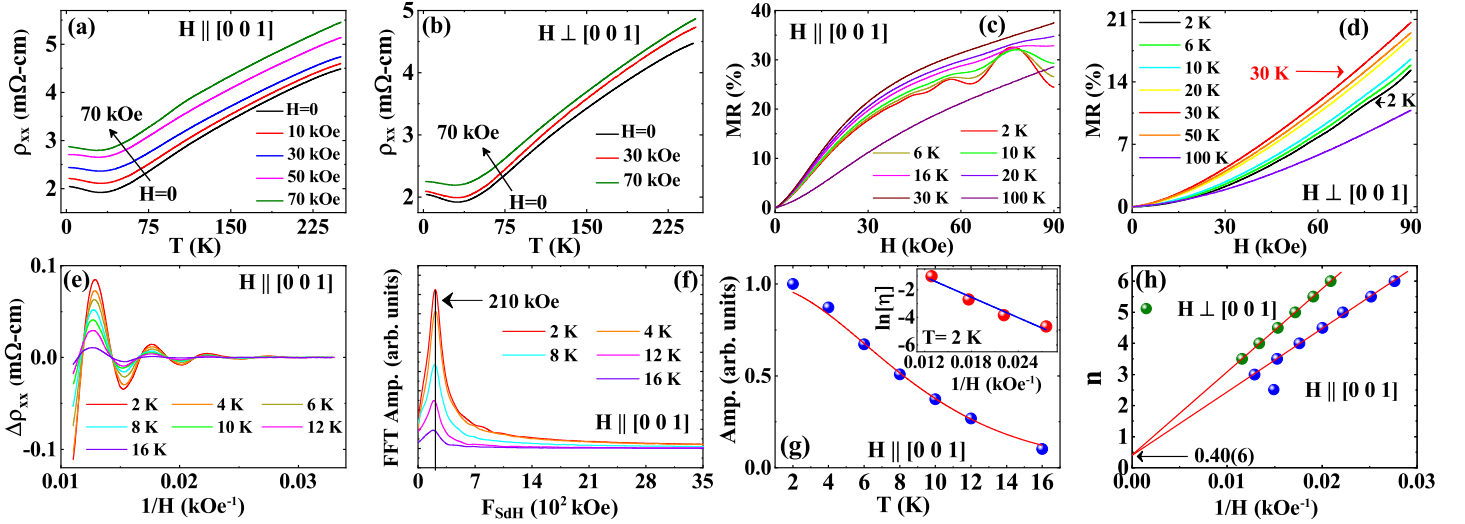}
\caption {$T$ variations of $\rho_{xx}$ at selected $H$ for $H$ (a) $||$ and (b) $\bot$ to [001] direction. $H$ dependence of MR (\%) for $H$ (c) $||$ and (d) $\bot$ to [001] direction at selected $T$. (e) Oscillatory component ($\Delta\rho_{xx}$) with 1/$H$ and (f) amplitude of FFT component with frequency ($F_{SdH}$) at selected $T$, showing a peak, as highlighted by the vertical line. (g) Oscillation amplitude with $T$. Inset of (g): Semi-log plot of $\ln[\eta]$ with 1/$H$ with $\eta = \rho_{xx}H\sinh(2\pi^{2}k_{B}T/\Delta E n(H))$. (h) Landau level fan diagram: $n$ with 1/$H$.} 
\label{MR}
\end{figure}

In this paper, we report bulk RSS in Bi$_2$Se$_3$ with dominant $p$-type conductivity, driven by hole doping. Synchrotron diffraction studies confirm structural transition from $R\overline{3}m$ to a non-centrosymmetric structure of $R$3$m$ around $\sim$ 30 K, below which significant negative thermal expansion is observed. First principles electronic structure calculations based on DFT for the $R3m$ structure confirm bulk RSS and the presence of surface state similar to the pristine Bi$_2$Se$_3$.
Transport experiment reveals Shubnikov-de Haas Van (SdH) oscillations in the low temperature magnetoresistance curves, where the Berry phase, as determined from the Landau level fan diagram, confirms the gap-less Dirac surface states along and $\bot$ to [001] direction. The saturation of magnetization curves at low temperature having a significantly large value of the magnetization suggest possible orbital magnetism. The result is consistent with the bulk RSS. The unique combination of orbital magnetism, bulk RSS, and conducting Dirac surface state in $p$-type (Bi$_{0.9}$Sb$_{0.1})_2$Se$_3$ systems is expected to be important for spintronic applications.

\section{Experimental and Computational Details}
Single crystal of (Bi$_{0.9}$Sb$_{0.1})_2$Se$_3$ is grown by using a modified Bridgman method\cite{tamal,ghose}. 
Synchrotron diffraction study is performed on the powdered crystal with a wavelength of 0.20694 \AA~ (60 keV) recorded at P02.1 beamline of PETRA III, Hamburg, Germany by using a Perkin Elmer XRD1621 area detector. The low temperature ($T$) studies are carried out by using a cryogen-free JANIS (USA) cryostat.

The Seebeck coefficient ($S$) and resistivity ($\rho$) are measured by using a home built set-up,\cite{tamal} coupled to a multifunctional probe of the PPMS system of Quantum Design (PPMS-II). The dc magnetization is measured by a SQUID VSM of Quantum Design. 
\par
First principles electronic structure calculations  within density functional theory (DFT) have been carried out using plane wave basis and projector augmented wave (PAW)\cite{PAW} potential as implemented in the Vienna Abinitio
Simulation Package (VASP)\cite{VASP}. For exchange correlation functional we have used Perdew-Burke-Ernzerhof (PBE) \cite{PBE} generalised gradient approximation (GGA). In the self consistent cycle, an energy cut off of 550 eV has been used with a gamma centered 9$\times$9$\times$3 k-mesh to execute the Brilliouin zone integrations. Low energy tight-binding model  Hamiltonian has been constructed using  maximally localized Wannier functions (MLWFs) following the formulation of Wannier90.\cite{Wannier90} The topological properties have been calculated using iterative Green's function approach\cite{green1,green2,green3,Wanniertools}.



\section{Results and discussion}
The quality of the crystal is checked by using X-ray diffraction studies and transmission electron microscopy of the powdered as well as single crystal sample (Fig.1 and Fig.2 of the SI \cite{supl}). 
The elemental analysis of the crystal using energy dispersive X-ray spectroscopy coupled with scanning electron microscope and X-ray photoemission spectroscopy confirms the desired composition, as discussed in the SI\cite{supl}. The
$T$ variations of $\rho_{xx}$ recorded in zero-field and magnetic field ($H$) applied $||$ and $\bot$ to [001] direction are shown in Figs. \ref{MR}(a) and \ref{MR}(b), respectively for different $H$. A minimum in $\rho_{xx} (T)$ is observed around $\sim$ 30 K ($T_m$). 
The magnetoresistance (MR), defined as [$\rho_{xx}(H) - \rho_{xx}(H=0)]/\rho_{xx}(H=0)$, are recorded, and depicted in Fig. \ref{MR}(c) for $H$ $||$ to [001] direction for selected $T$. A clear signature of oscillation is observed at 2 K, where maxima as well as minima in $\rho_{xx}(H)$ are found at a regular interval of 1/$H$, pointing to a typical manifestation of the SdH oscillation\cite{kra,mandal1,lv1}. The SdH oscillation is evident only at 2 K for the $\bot$ component, having a much smaller amplitude and lower frequency, as shown in Fig. \ref{MR}(d). The values of MR(\%) increase with $T$ for both the components, and then, they decrease with further increasing $T$, showing a maximum around $T_m$.

Figure \ref{MR}(e) shows the periodical oscillatory component ($\Delta\rho_{xx}$) with $1/H$ for the $||$ component. The $\Delta\rho_{xx}$ is obtained by subtracting the background from the $\rho_{xx} (H)$ isotherms. The Fast Fourier transform (FFT) of $\Delta\rho_{xx}(H)$ is depicted in Fig. \ref{MR}(f), which  exhibits a peak around 210 kOe for the measurement along [001], analogous to the observed single peak for the pristine compound\cite{eto_BS,busch_BS,dev_BS}. 

A single peak is also observed around 2670 kOe for $H$ $\bot$ to [001] direction.
The oscillation frequency provides the extremal area of cross section ($A_F$) of the Fermi surface (FS) according to the Onsager relation, $F_{SdH}=(\hbar/2\pi e)A_F$, with $A_F=\pi k_F^2$. The difference between the $||$ and $\bot$ component in $F_{SdH}$ points to a highly anisotropic FS, analogous to Bi$_2$Se$_3$\cite{eto_BS,qu_BS}. The values of $F_{SdH}$ provide the values of Fermi momentum ($k_F$) $\sim$ 2.5 $\times 10^{-2}$ and 9.0 $\times 10^{-2}$ $\AA^{-1}$ for the $||$ and $\bot$ components, respectively. 

Temperature dependence of the amplitude of the oscillations at 210 kOe follows the standard Lifshitz-Kosevich (L-K) expression,\cite{ando_LK,shoe} as shown in Fig. \ref{MR}(g) by the continuous curve. 
The value of the effective cyclotron mass ($m^{\ast}$), as obtained from the fit, is 0.14 $m_e$, where $m_e$ is the free electron mass. Thus, the Fermi velocity is obtained as 2.08 $\times 10^{5}$ m/s using $v_F = \hbar k_F/m^{\ast}$, which provides the value of Fermi energy, $E_F \approx \frac{1}{2}m^{\ast}v_F^2 \approx$ 17.34 meV. The slope of the semi-log plot of $\ln[\rho_{xx}H\sinh(2\pi^{2}k_{B}T/\Delta E_{n}(H))]$ with $1/H$ provides the value of Dingle temperature, $T_D$ = 11.99 K, as shown in the inset of Fig. \ref{MR}(g). The life time ($\tau$) of the surface charge carriers is obtained to be 1.01 $\times$ 10$^{-13}$ s using $\tau = \hbar/(2\pi k_B T_D$), which provides the mean free path, $l = \tau v_F$ and mobility, $\mu = e\tau/m^{\ast}$ as 21.14 nm, and 0.80 $\times 10^{4}$ cm$^2$/V s, respectively.
These values are close to the values for Bi$_2$Se$_3$\cite{eto_BS,busch_BS}.
The $n$th maxima observed in the MR-$H$ curves are plotted with 1/$H$ in Fig. \ref{MR}(h). The extrapolation of the linear plots for both the $||$ and $\bot$ component, the representative of the Landau level fan diagram, meet at $1/H$ = 0 with $n$ = 0.40(6), which is close to 0.5 recommended for the Dirac particles,\cite{shrestha1,shrestha2,mandal1} and consistent with the results of pristine Bi$_2$Se$_3$\cite{eto_BS}. Our results suggest that the topologically nontrivial states are retained even for nearly 10 \% Sb doping in Bi$_2$Se$_3$.

\begin{figure}
\centering
\includegraphics[width =0.7\linewidth]{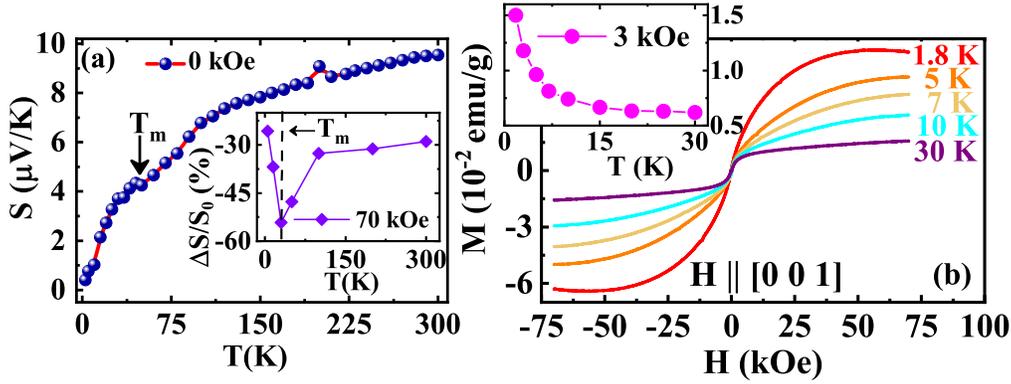}
\caption {(a) Thermal variations of $S$. Inset of (a): $\Delta S/S_0$ with $T$ at 70 kOe applied along [001] direction. (b) Magnetization curves ($M$ vs. $H$) at selected $T$ for $H$ applied along [001] direction. Inset depicts the plot of $M$ with $T$ at 30 Oe.} 
\label{See}
\end{figure}

The carrier concentration and mobility are investigated from the Hall measurement, which is given in Fig.3 of the SI\cite{supl}. The Hall results confirm the dominant $p$-type conduction. 
The values of Seebeck coefficient $S$ with $T$ are shown in Fig. \ref{See}(a). At room temperature, the value of $S$ is positive, indicating a dominant $p$-type conduction. The $p$-type conductivity has also been realized by the Ca\cite{hor_PRB} and Mn\cite{choi_APL} doping in Bi$_2$Se$_3$.
The value of $S$ decreases with decreasing $T$ down to 2 K, displaying a shoulder at $T_m$ as indicated by an arrow\cite{takaga}. The $H$ dependence of $S$ is recorded at selected $T$, are shown in Fig.4 of SI,\cite{supl} where $H$ is applied along [001] direction. 
The value of $\Delta S/S_0$, defined as [$S(H) - S(H = 0)]/S(H = 0)$, is plotted with $T$ at $H$ = 70 kOe,  in the inset of Fig. \ref{See}(a). The value decreases with decreasing $T$ and exhibits a minimum close to $T_m$, below which it increases sharply. The maximum value of $|\Delta S/S_0|$ is $\sim$ 55 \% for $H$ = 70 kOe. The characteristic $T$ variations of $S$ and magneto-Seebeck response indicate
the possibility of RSS below $T_m$, as suggested in the theoretical works\cite{is_TE,ver_TE,wu_jing1}.

Figure \ref{See}(b) displays the magnetization ($M$) curves at selected $T$ in the low-$T$ region with $H ||$ to [001] direction. In the inset the value of $M$ at 3 kOe shows a paramagnetic-like increase with decreasing $T$ below $T_m$. Any detectable coercivity is absent in the magnetization curves. The curves exhibit a saturating trend of magnetization at high-$H$, except for the curve at 1.8 K. A decreasing trend of $M$ is observed for high-$H$, indicating a diamagnetic component. This is consistent with the Berry phase component, which is anticipated to exist at high-$H$ end of the magnetization curves. The saturating trend of magnetization without coercivity is significant, which is similar to that observed for Bi-rich Bi$_2$Se$_3$ nanoplatelets, pointing to a possible manifestation of the orbital magnetism\cite{kim_BS5}. Although a handful of the theoretical studies have been performed on orbital magnetism,\cite{xiao_orbit,lux_orbit} the experimental verification of it are rare\cite{sch_orbit}. We note that the value of saturation of magnetization is $\sim 6.5 \times 10^{-2}$ emu/g at 1.8 K, which is considerably high. The origin of the orbital magnetism in topological insulators may be traced back to Rashba spin splitting as argued by Kim $et.$ $al$.\cite{kim_BS5} In order to understand the origin of RSS in centrosymmetric Bi$_2$Se$_3$, we carried out low-$T$ synchrotron diffraction studies to identify any structural transition upon Sb doping at the Bi site, that may promote hosting of the bulk RSS.   

\begin{figure}
\centering
\includegraphics[width =0.8\linewidth]{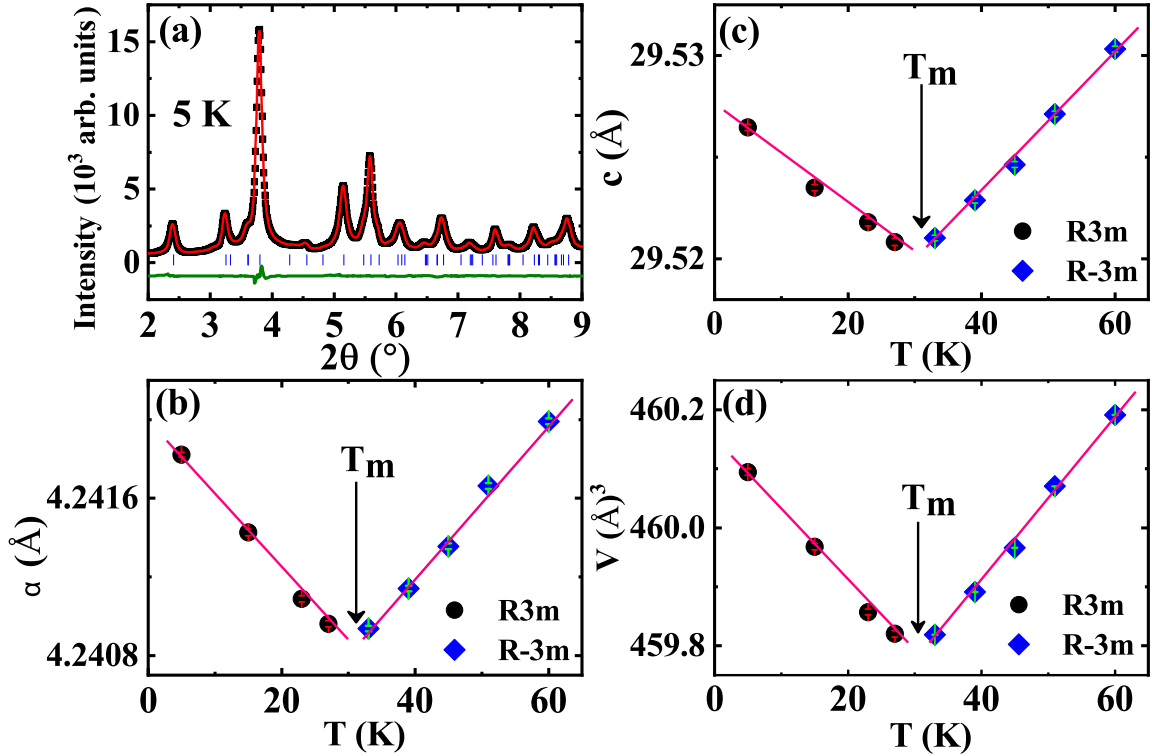}
\caption {(a) Synchrotron diffraction pattern with the Rietveld refinement, as indicated by the continuous curve at 5 K. Thermal variations of (b) $a$, (c) $c$, and (d) $V$.} 
\label{LTXRD}
\end{figure}

Synchrotron diffraction studies are performed at low-$T$ in the range of 5-60 K. As depicted in  Fig.5(a) of the SI\cite{supl}, the integrated intensity of a diffraction peak exhibits a significant change around $T_m$, indicating a possible structural change\cite{jie_shen}. 
The ISODISTORT \cite{isodistort} software is used to search possible structural phases below the transition. We note that the $R3m$ space group has the highest symmetry within the possible structures. A comparison of the refinement at 5 K using $R3m$ (160) and high-$T$ $R\overline{3}m$ (166) structure is shown in Figs.5(b) and 5(c) of the SI,\cite{supl} validating satisfactory refinement with $R3m$ space group, which is a non-centrosymmetric structure. Figure \ref{LTXRD}(a) depicts the refinement of the synchrotron diffraction pattern at 5 K using $R$3$m$ space group. The refined coordinates are Bi1/Sb1 (0, 0, 0.3993(3)), Bi2/Sb2 (0, 0, 0.6006(7)), Se1 (0 0 0), Se2 (0 0 0.2101(3)), and Se3 (0 0 0.7898(7)) with the reliability parameters, R$_w$ (\%) = 5.15, R$_{exp}$ (\%) = 2.20, and $\chi^2$ = 2.33. The lattice constants, $a$, $c$, and the unit cell volume ($V$) are shown in Figs. \ref{LTXRD}(b), \ref{LTXRD}(c), and \ref{LTXRD}(d), respectively. The results indicate a structural transition close to $T_m$.  Below $T_m$, the negative thermal expansion in $V(T)$ is observed, which is $\sim$ 0.07 \% at 5 K with respect to the value at $T_m$. This structural transition at $T_m$ may be correlated with the observed electrical transport and Seebeck coefficient. As shown in Figs. \ref{MR}(a) and \ref{MR}(b), a minimum is observed in the MR curves, whereas an anomaly in $S(T)$ is highlighted by an arrow in Fig. \ref{See}(a) close to the structural transition, at $T_m$.

\begin{figure} 
\includegraphics[width=\linewidth]{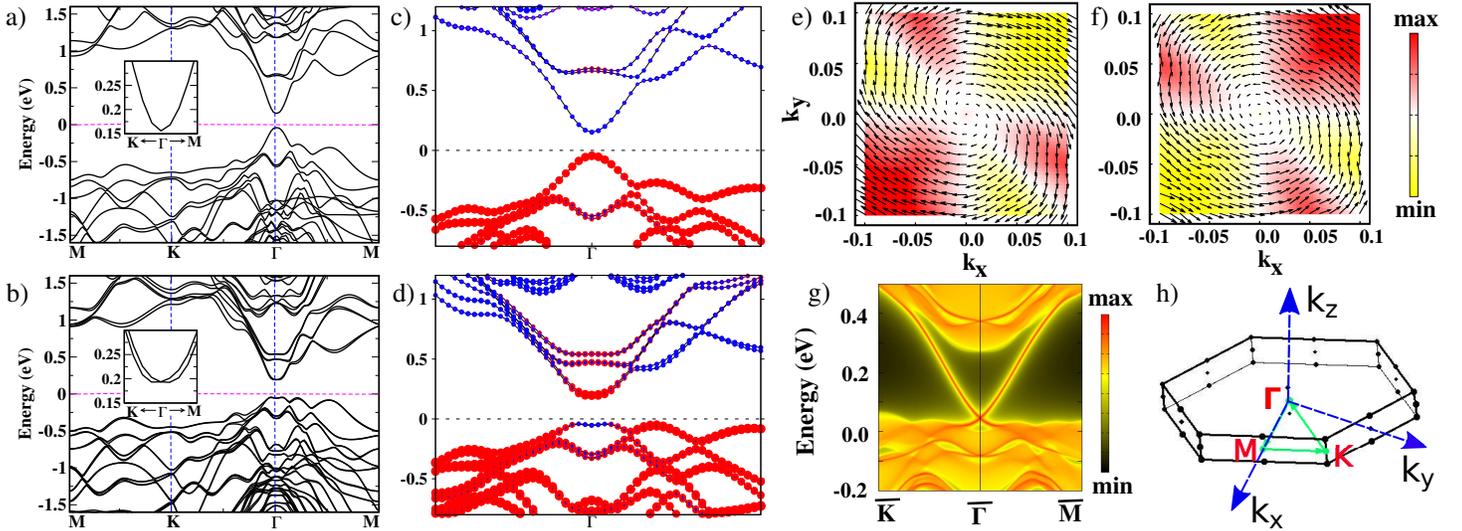}
\caption{(a,b)The electronic band structure of Bi$_{2-x}$Sb$_x$Se$_3$ without and with SOC respectively. The inset shows enlarged view of the conduction bands around the $\Gamma$ point. (c,d) The orbital characters projected on the band structure without and with SOC respectively. Red, blue and magenta color indicates  Se-$p$, Bi-$p$ and Sb-$p$ states. Fermi energy is set to zero in the energy axis. (e-f) The spin textures of the inner and outer branch of the conduction band around the $\Gamma$ point. Color bar indicates the contribution of the out of plane spin component. (g) Surface state spectrum along [001] direction, here color bar indicates the contribution of the bands. Black color indicates the bulk energy gap. (h) Brilliouin zone of the system.}
\label{band}
\end{figure}
Our experimental results discussed above not only suggests the presence of Dirac surface state but also predicts the possibility of bulk RSS. In view of this, electronic structure calculations have been carried out using the experimental lattice parameters of the low-$T$ $R3m$ structure. In order to simulate doping, we have substituted one of the six Bi sites with a Sb atom in the unit cell. This leads to  Bi$_{2-x}$Sb$_x$Se$_3$ with $x=0.16$ which is adequate to capture the experimental doping concentration of 10$\%$ (i.e. $x=0.2$). Subsequently,  doping breaks the inversion symmetry of the system. The non spin polarised band structure has been shown in the Fig. \ref{band}(a). The directions of the k-path and the high symmetry k-points, which are used to calculate the 
band structure, have been shown in Fig. \ref{band}(h). The bulk  band structure indicates that the system is an insulator with a calculated direct band gap of 0.2 eV. The orbital projected non spin polarised band structure has been shown in Fig. \ref{band}(c) to identify the orbital contribution in the low energy valence and conduction bands. The occupied Se-$p$ states lie from -4 eV up to the Fermi level to form the low energy valence bands. While, the conduction bands are mainly formed with completely unoccupied Bi-$p$ and Sb-$p$ states.
\par
The SOC included band structure has been shown in Fig. \ref{band}(b), which clearly shows that, inclusion of SOC lifts the degeneracy of the bands and splits them along some particular directions in the k-space. In particular, we have considered the low lying conduction bands around the $\Gamma$ point and shown the enlarged view of the band structures without and with SOC in the inset of Fig. \ref{band}(a) and \ref{band}(b) respectively. A substantial amount of splitting of the bands are found along $\Gamma$-M and $\Gamma$-K directions of the Brillouin zone, perpendicular to the (0001) direction, which  indicate the existence of Rashba-Dresselhaus effect in the system. In Fig. \ref{band}(e) and \ref{band}(f) we have shown the spin textures of inner and outer branch of the spin split bands around the $\Gamma$ point. The rotating nature of the in plane spin components around the $\Gamma$ point indicates the dominance of the Rashba effect. The finite value of the out of plane spin component may be attributed due to the presence of the higher order non-linear terms\cite{NLHO}. The system belongs to the C$_{3v}$ point group symmetry, which can host out of plane spin component through the higher order Dresselhaus term, which has been proposed recently\cite{LWN}. Further, from the band splitting we have estimated the linear Rashba ($\alpha_R$) and Dresselhaus ($\alpha_D$) parameters of the system around the $\Gamma$ point.\cite{Tsymbal} Our calculated values of $\alpha_R$ and $\alpha_D$ are 0.6 and -0.2 eV\AA,  respectively, that reveals predominant contribution of the  Rashba effect around the $\Gamma$ point. The presence of Rashba effect induces an in-plane $k$-dependent relativistic magnetic field, and is expected to manifest in the transport properties.

The fatness of the bands in the presence of SOC (see Fig. \ref{band}(d)) shows the presence of Bi-$p$ character in the valence band maxima whereas, Se-$p$ character in the conduction band minima at the $\Gamma$ point. This implies an inversion between Bi-$p$ and Se-$p$ states in presence of SOC, suggesting the possibility of hosting topological properties. To explore the topological properties we have calculated the surface band structure of a semi infinite slab of Bi$_{2-x}$Sb$_x$Se$_3$, which is constructed in a rotated frame having the quintuple layers along 0001 direction. Surface band structure is then projected into the bulk band structure and the result is shown in Fig. \ref{band}(g). Surface band structure show conducting states which form a Dirac cone in the bulk energy gap, which is the signature of a topological insulator. Topological systems that lack inversion symmetry are of particular interest as they can  simultaneously host conducting state at the surface and Rashba effect in the bulk, enhancing their scope of  application.

To conclude, the proposed RSS elegantly construes the correlation between structural, electrical, magnetic, and thermoelectric transport results. Below $T_m$, occurrence of bulk RSS is driven by the combined effect of structural distortion to a non-centrosymmetric structure and the strong SOC. Our experimental results of bulk RSS and associated nontrivial conducting surface state in the $p$-type Sb doped Bi$_2$Se$_3$, is further corroborated with first principles electronic structure calculation.  



\medskip
\textbf{Supporting Information} \par 
Supporting Information is available from the Wiley Online Library or from the author.

\medskip
\textbf{Acknowledgements} \par 
Synchrotron diffraction studies were performed at the light source PETRA III of DESY, a member of the Helmholtz Association (HGF). Financial support (Proposal No. I-20180276) by the Department of Science \& Technology (Government of India) provided within the framework of the India@DESY collaboration is gratefully acknowledged.  SG acknowledges DST (Project No. SB/S2/CMP029/2014) for the instrumental facility used in this study. ID thanks Technical Research centre DST and SERB (Project No. EMR/2016/005925) for financial support.

\medskip

%







\


\end{document}